\documentclass[11pt]{article}

\usepackage[T1]{fontenc}
\usepackage[utf8]{inputenc}
\usepackage{lmodern}
\usepackage{amsmath}
\usepackage{amssymb}
\usepackage[margin=1in]{geometry}
\usepackage[round,authoryear]{natbib}
\usepackage{booktabs}
\usepackage{tabularx}
\usepackage{longtable}
\usepackage{array}
\usepackage{graphicx}
\usepackage{placeins}
\usepackage[dvipsnames]{xcolor}
\PassOptionsToPackage{hyphens}{url}
\usepackage{hyperref}

\definecolor{LinkBlue}{HTML}{1F4E79}
\definecolor{UrlBlue}{HTML}{005A9C}
\definecolor{ReferenceGreen}{HTML}{21613E}

\hypersetup{
  colorlinks=true,
  linkcolor=LinkBlue,
  citecolor=UrlBlue,
  urlcolor=UrlBlue,
  breaklinks=true,
  pdfborder={0 0 0},
  pdftitle={When Statisticians Can Also Be AI Engineers, What Could We Build?},
  pdfsubject={Agentic AI and the Distribution of Statistical Inquiry}
}
\newcolumntype{L}[1]{>{\raggedright\arraybackslash}p{#1}}

\newif\ifPaperMainCitations
\PaperMainCitationstrue
\newif\ifPaperPrintMainRefs
\makeatletter
\let\PaperOriginalSortCites\NAT@sort@cites
\renewcommand{\NAT@sort@cites}[1]{%
  \PaperOriginalSortCites{#1}%
  \ifPaperMainCitations
    \@for\PaperCiteKey:=\NAT@cite@list\do{%
      \expandafter\gdef\csname PaperMainRef@\PaperCiteKey\endcsname{}%
    }%
  \fi
}
\makeatother
\long\def\PaperGobbleEntry#1\par{\endgroup}
\def\PaperDiscardEntry{\begingroup\catcode`\%=12\catcode`\^=12\PaperGobbleEntry}
\def\PaperFilterBibitem[#1]#2{%
  \let\PaperNextBibitem\PaperDiscardEntry
  \ifcsname PaperMainRef@#2\endcsname
    \ifPaperPrintMainRefs
      \def\PaperNextBibitem{\PaperOriginalBibitem[#1]{#2}}%
    \fi
  \else
    \ifPaperPrintMainRefs\else
      \def\PaperNextBibitem{\PaperOriginalBibitem[#1]{#2}}%
    \fi
  \fi
  \PaperNextBibitem
}
\newenvironment{paperreferences}[1]{%
  \raggedright
  \renewcommand{\refname}{#1}%
  \let\PaperOriginalBibitem\bibitem
  \let\bibitem\PaperFilterBibitem
}{}

\title{\textbf{When Statisticians Can Also Be AI Engineers, What Could We Build?}\\[0.5em]
\large Agentic AI and the Distribution of Statistical Inquiry}
\author{Ruoqing Zhu\\[0.3em]
\small Department of Statistics\\
\small University of Illinois Urbana-Champaign\\
\small \href{mailto:rqzhu@illinois.edu}{\texttt{rqzhu@illinois.edu}}}
\date{}

\begin{document}
\maketitle

\begin{abstract}
Statistical software makes methods widely available. Good statistical work, however, requires understanding the question, how the data were collected, and what conclusions the evidence can support. As AI agents make data analysis accessible to people without statistical training, this reasoning should be built into the systems themselves. Our review of selected AI agents for data analysis and study design points to a gap in documented support for scientific inquiry with users at runtime. In this paper, we advocate designing domain-specific interactive harnesses in which statistical principles guide proactive inquiry: agents recognize missing information that could change an analysis, ask questions users may not know to raise, and use the answers to guide or revise analytical decisions. Two practical entry points illustrate how statisticians can begin building: an implemented causal-consulting skill within an existing agent environment and Model Forge, a custom harness for interactive model and algorithm development. These design examples connect statistical principles to choices about context, memory, tools, and interfaces, and motivate evaluation of both computational correctness and the scientific consequences of interaction. We conclude with a broader interdisciplinary research agenda on adaptive human-agent interaction, including selective information acquisition, evolving user preferences, and evaluation under uncertainty.

\end{abstract}

\section{Introduction}
\label{sec:introduction}

For decades, statisticians have used software to disseminate statistical knowledge. A method gains far greater reach when researchers can install a package, apply it to their own data, inspect examples, and reproduce an analysis. S and R made this form of dissemination central to modern practice by providing extensible environments for methodological research, collaboration, and distribution \citep{ihaka1996,chambers2020}. Packages, documentation, and reproducible research compendia consequently became scholarly objects that carry algorithms, defaults, examples, and workflows far beyond their creators \citep{gentleman2007}. Access to statistical methods does not by itself resolve the choices connecting a scientific question to an analysis: how to define the problem, which information is relevant, and what conclusions the available data can support. Statistical consultation treats these choices as provisional. The statistician asks whom the result should describe, how observations entered the study, what was measured and when, what is missing, which comparison matters, and what decision the analysis will inform. Classic accounts therefore describe consultation as an iterative process of identifying the problem situation, defining goals, deciding on actions, and clarifying responsibilities \citep{zahn1983}. Mallows called the prior question of whether the data are relevant to the substantive problem the ``zeroth problem'' of statistical analysis \citep{mallows1998}.

These questions determine whether a calculation has evidential meaning. Study design may matter more than subsequent model fitting, especially for causal inference \citep{rubin2008}, and large samples cannot repair systematic defects in how observations entered a dataset \citep{meng2018}. The American Statistical Association's ethical guidelines accordingly include data collection, analysis, interpretation, and model deployment within statistical practice, and call on practitioners to understand stakeholder goals, communicate provenance and assumptions, and consider how an analysis may fail in context \citep{asa2022}. Sosa and Betancourt develop these obligations as \emph{statistical warrant}: the target, observation regime, assumptions, procedure, uncertainty assessment, validation criterion, consequences, and accountability connecting data to a claim \citep{sosa2026}. A calculation can therefore be correct under a model while failing to support the conclusion for which it is used. Software that helps establish and revisit this connection could distribute a substantial part of statistical practice.

Agentic AI creates an opportunity to bring this reasoning into the course of an analysis. An agent can inspect files, write and execute code, examine results, and decide what to do next. Prominent data-agent systems organize these capabilities around completing a requested task, such as preparing data, fitting a model, or producing a report \citep{sun2025lambda,hong2024}. This makes automation a natural starting point for understanding their value: a user supplies an objective and materials, and the agent coordinates work toward an output. The same capabilities also raise an engineering question. What else could the system be designed to do as the work unfolds?

Someone who has never analyzed data might ask an AI agent, ``Do I sleep longer on days when I walk more?'' An agent can compare daily step counts with sleep the following night \citep{merrill2026wearable}. But suppose the apparent pattern depends on a handful of days with very few recorded steps. A useful next question would be, ``Were those quiet days, or were there times when you took your watch off?'' If the person removes the watch at work, a low count has a different meaning. The agent could show whether the pattern persists on days with more complete records and explain which days the comparison describes. The person knows when they wear the watch. They should not have to recognize a measurement problem before the agent thinks to ask. Here statisticians have both an opportunity and a responsibility: they can help design what an agent notices, what it asks, and what conclusions it draws from the answer.

We use \emph{harness} for the instructions, software, tools, context and memory, interfaces, and execution controls surrounding a large language model (LLM). These components govern what information the model receives, which actions it can take, what persists, and how the user participates \citep{bonamy2026}. An \emph{agent} is the LLM operating through this harness. A builder can organize the harness to support interactive exploration. Alongside examining data and consulting background materials, the agent can actively explore the user's goals and domain knowledge, including facts the user did not know were relevant to the analysis. For example, apparent repeated measurements may prompt a question about whether identifiers denote the same people over time. The answer could support the proposed analysis, require a different representation of the data, or reveal that further documentation is needed. The system's task thus includes developing an understanding of the problem while it works toward a result.

We use \emph{domain-specific interactive harness} for a harness that organizes an agent's work around a domain's goals, reasoning practices, and standards of evidence, with user participation contributing to how the problem and its solution develop. Its \emph{policy for inquiry} governs what to inspect, when to seek user input, how to interpret that input, and how it should inform subsequent work. Through this interaction, the problem formulation, relevant information, and direction of work can evolve. This paper examines how statistical principles can shape such harnesses. Understanding the observation process can guide what the agent inspects and asks; distinguishing evidence from assumptions can guide its memory; and concern for justified conclusions can guide when it revises a plan or limits a claim. The harness organizes this behavior around the LLM, while the quality of the resulting reasoning also depends on the model. This direction builds on earlier interactive statistical software and recent statistical advocacy, developed more concretely in Section~\ref{sec:runtime-interaction} \citep{pregibon1991,lin2026lambda}.

Statistical engineering, as developed by \citet{hoerl2017statisticalengineering,hoerl2020statisticalthinking}, emphasizes integrating statistical methods, information technology, and domain knowledge to address complex, initially unstructured problems. The present paper shares this concern with translating statistical expertise into effective practice and examines a further design opportunity: building reusable AI systems that participate in statistical inquiry through interaction with their users. How these systems ask questions, interpret answers, retain scientific context, and revise their work becomes an object of statistical design and evaluation, with applications spanning consultation, methodological research, and education.

This paper focuses on one part of AI engineering: designing and evaluating agents built around existing language models. We develop two entry points with different degrees of customization: adapting an existing harness through the interactive \emph{causal-consultant} skill, and assembling a custom interactive harness from reusable engineering infrastructure, illustrated by Model Forge. Coding agents can assist with implementation, while the statistician specifies the intended behavior and evaluates its statistical consequences. The practical aim is to make a first attempt possible and to give its author a concrete way to examine what the interaction contributes.

The discussion spans consultation, collaborative model development, theoretical research, and teaching, with the user's expertise determining the division of work. We trace the historical and statistical rationale for interaction, introduce skills through a consultation example, and review existing data analysis agents. Model Forge then illustrates custom harness development, followed by reusable infrastructure, engineering practices, and evaluation. The final section considers how agents can learn about users while helping them apply statistical principles.

\section{A Historical Perspective on Technology and Statistical Practice}
\label{sec:technology-history}

The history of statistics repeatedly shows that new technologies do more than accelerate existing tasks. They change which questions are practical, what counts as a statistical method, and how statistical knowledge travels. Electronic computing made Monte Carlo methods and repeated simulation central to research on problems that were difficult to attack analytically \citep{metropolis1987,metropolisulam1949}. The bootstrap went further by making computation part of the inferential method itself \citep{efron1979bootstrap}. Efron described this broader shift as allowing statistical theory to begin ``thinking the unthinkable'' \citep{efron1979computers}. The important point is the pattern: a new technical substrate changed both how statisticians reasoned and which forms of reasoning could be put into practice.

Statistical software produced a related change in dissemination. S and R provided extensible environments for data analysis, while packages and reproducible artifacts allowed methods to circulate with executable code, defaults, examples, and documentation \citep{chambers2020,ihaka1996,gentleman2007}. Shiny extended this progression by letting statisticians distribute interactive applications to people who did not write code \citep{cheng2012shiny}. The recognition of John Chambers for S and Hadley Wickham for R-based computing, visualization, and accessibility reflects how software infrastructure itself became influential statistical work \citep{asa1998chambers,copss2019wickham}. Across these developments, the unit of knowledge that could travel expanded from an explanation of a method, to an executable procedure, to an interactive space for exploring an analysis.

Distributing statistical judgment introduces a further design problem: how can software obtain the information that makes a method appropriate, and how can the user understand and revise the resulting analysis? Earlier expert systems and interactive analytical tools addressed parts of this problem. Section~\ref{sec:runtime-interaction} examines their statistical contributions alongside the recent calls for scientific dialogue in agentic systems.

Agentic AI changes the practical means of addressing this problem. An agent can inspect local materials, communicate in natural language, call computational tools, and respond to information discovered while the work is underway. Coding agents can also help a statistician implement and revise the harness. Existing statistical packages supply analytical methods, while reusable agent runtimes supply much of the conversation and execution infrastructure \citep{bonamy2026}. These capabilities expand what a small team or individual can attempt. Together, coding assistance and reusable agent infrastructure give statisticians a more accessible route into AI engineering: they can implement and examine systems whose behavior reflects their statistical ideas. Specifying useful behavior and evaluating the resulting interaction remain substantial parts of this work. The question is therefore both what to build and how to begin: which part of statistical practice could become interactive, what can be reused, and which decisions require the builder's statistical judgment?

\section{Interactive Statistical Reasoning at Runtime}
\label{sec:runtime-interaction}

The engineering view raises a statistical question: what should a system learn about a problem while an analysis is taking shape, and how should that knowledge change its work? Interactive systems offer one concrete setting in which statisticians can specify the reasoning they want software to support.

For consultation, an especially useful goal is to support a domain scientist who knows how treatment was assigned, why measurements are missing, whether observations came from the same person or institution, and what decision the result will inform, but has little statistical training. The system would elicit this knowledge in ordinary language and take responsibility for constructing a provisional statistical representation. This is an accessibility objective to design and evaluate, rather than a capability that follows automatically from using a language model. In theoretical collaboration, the user may instead be a statistical expert; in education, developing the user's statistical understanding may be the purpose of the interaction.

The resulting questions should concern the world the user knows. Instead of asking whether observations are independent, a consultation system might ask whether the same person can appear more than once or whether several people came from the same clinic. Instead of asking whether conditional exchangeability is plausible, it might ask how treatment was assigned and whether that decision depended on information absent from the dataset. The answers provide evidence relevant to dependence or identification; they do not automatically establish the corresponding assumptions. A useful system would explain the consequences, preserve unresolved alternatives, and seek clarification when the statistical interpretation remains uncertain.

The same interaction can support developing a new model or algorithm, beginning from a methodological question, a proposed formulation, or a practitioner's data. For example, a collaborator might explain that measurements were taken more frequently when participants' conditions worsened. That information could motivate a model representing both the outcome trajectory and the observation process. The agent could help formulate candidate models, examine which parameters the available data can identify, and design simulations to compare their behavior. The interaction would then support methodological development as well as assessment of an existing analysis. Such a dialogue could connect model choice, assumptions, theoretical validity, computation, and scientific usefulness. It should distinguish a property proved under stated assumptions from favorable simulation results or a good fit to the observed data.

When information needed to define a target or assess a design is absent from the data, more computation may not resolve the uncertainty. Relevant knowledge may reside with investigators, data collectors, domain experts, or affected stakeholders; sometimes no available participant knows the answer. Asking, requesting documentation, revising a representation, narrowing a claim, and seeking specialist input can therefore be legitimate analytical actions. The design task is to connect them: an answer should alter the account of the problem and the subsequent work when its statistical consequences require a change. Success includes recognizing what remains unknown.

The effort to put such reasoning into software has a substantial history. Statistical expert systems examined how analytical software might select procedures, request information, and explain recommendations \citep{hand1984,pregibon1991}. Mixed-initiative interfaces studied when a system should act or seek user input, accounting for uncertainty and the cost of interruption \citep{horvitz1999mixedinitiative}. More recent statistical tools make the connection to problem formulation concrete. Tisane derives model structure from conceptual and measurement relationships, including statistical components that a user need not know how to specify. Its successor, rTisane, supports externalizing implicit domain assumptions and resolving conceptual ambiguity \citep{jun2022tisane,jun2024rtisane}. StatZ supplies conversational guidance, GuidedStats supports revisable workflows and assumption checks, and CausalChat helps analysts refine causal graphs \citep{wen2025statz,zhang2024guidedstats,zhang2024causalchat}. These systems offer working examples and design lessons, with different supported models and expectations of the user.

Recent statistical commentary articulates why this interaction matters and what it should accomplish. In her discussion of LAMBDA, Lin proposes analysis-planning and result agents that engage users in scientific dialogue, refine goals, flag statistical pitfalls, and identify matters requiring user input \citep{lin2026lambda}. Meng's proposal for data minding emphasizes prompting users to reflect on data provenance and quality \citep{meng2026mindware}. Zou and Yuksekgonul advocate agents that seek clarification and collaborate with users, while Liu and colleagues discuss learning statistical reasoning through analytical interactions \citep{zou2026lambda,liu2026lambda}. Together, these discussions supply a statistical rationale and design agenda: scientific framing, assumptions, and interpretation deserve attention throughout the interaction. Translating that agenda into a product requires further choices about how questions are triggered, what answers mean, which information persists, and how the behavior is evaluated. The two entry points developed below address these implementation decisions.

In statistical applications, a domain-specific interactive harness can distribute such a policy for inquiry. It connects that policy to tools and information, preserves decisions, and presents the work to the user. We advocate building and examining such harnesses across statistical practice, using existing infrastructure wherever it serves the task. A first attempt need not include a custom interface or a new agent runtime. It can begin by expressing the intended interaction as a skill loaded within an existing harness, then observing whether different user answers produce the intended statistical consequences.

\section{A First Entry Point: Adapting a Harness with a Skill}
\label{sec:skills}

\subsection{Skills as Executable Task Packages}
\label{subsec:skills-history}

A statistician can begin developing a domain-specific interactive harness by adapting an existing agent environment with a skill. A useful first project is to make one statistical idea interactive: identify when it matters, what information its application requires, and how a user's answer should change the work. An agent skill provides an accessible way to express this policy. The basic role of skills is to make task-specific context and instructions reusable and available on demand. A skill is a portable package, usually organized around a \texttt{SKILL.md} file, that supplies these instructions together with optional scripts, references, templates, and data assets. The host loads the relevant instructions and makes supporting resources available as needed \citep{agentskills2025}. This allows the author to specify what to inspect, which questions to ask, and how answers should revise an analysis within an existing agent environment.

Skills have already grown well beyond this basic role. The format supports scripts, and current skills can organize substantial executable machinery. Supporting scripts can perform specialized operations, check inputs and outputs, maintain persistent records, or provide access to a database. Instructions can define the intended scope of work, while code checks enforce particular operational conditions. The skill remains a component of the harness: the host supplies model access, execution capabilities, permissions, and the surrounding interface. Its complexity can therefore extend well beyond the text placed in the model's context.

Archify and Anatomy illustrate this broader design space. Archify combines diagram-authoring instructions with structured schemas, rendering code, an interactive viewer, and automated validation \citep{tta1i2026archify}. Anatomy combines illustration guidance with a geometry library, SVG and WebGL components, and scripts for geometric and browser-based checks \citep{ryan2026anatomy}. Their instructions organize a workflow supported by reusable software. A statistical skill can use the same approach to connect a policy for inquiry to computation, persistent scientific records, and checks on how the work proceeds.

\subsection{A Causal-Consulting Conversation Loop}
\label{subsec:skills-consultation}

Causal inference gives a concrete starting point. An estimator becomes meaningful only in relation to a target and a defensible account of how observations and treatment arose \citep{rubin2008}. Consultation develops that account through questions, inspection, explanation, and revision \citep{zahn1983,pregibon1991}. The author's \emph{causal-consultant} skill, examined here in its public version 8.0.7, packages this process with scientific role guides and a supporting memory utility \citep{zhu2026causalconsultant807}. We use this implementation to illustrate how a statistical consultation policy can be expressed through an interactive skill. The discussion describes its structure and intended behavior; its scientific effectiveness, which also depends substantially on the particular LLM used for inference, is not evaluated here. It can support planning a new study, examining existing data, or interpreting evidence; the next contribution depends on the user's question and what the current record supports.

\begin{figure}[!t]
\centering
\includegraphics[width=\textwidth]{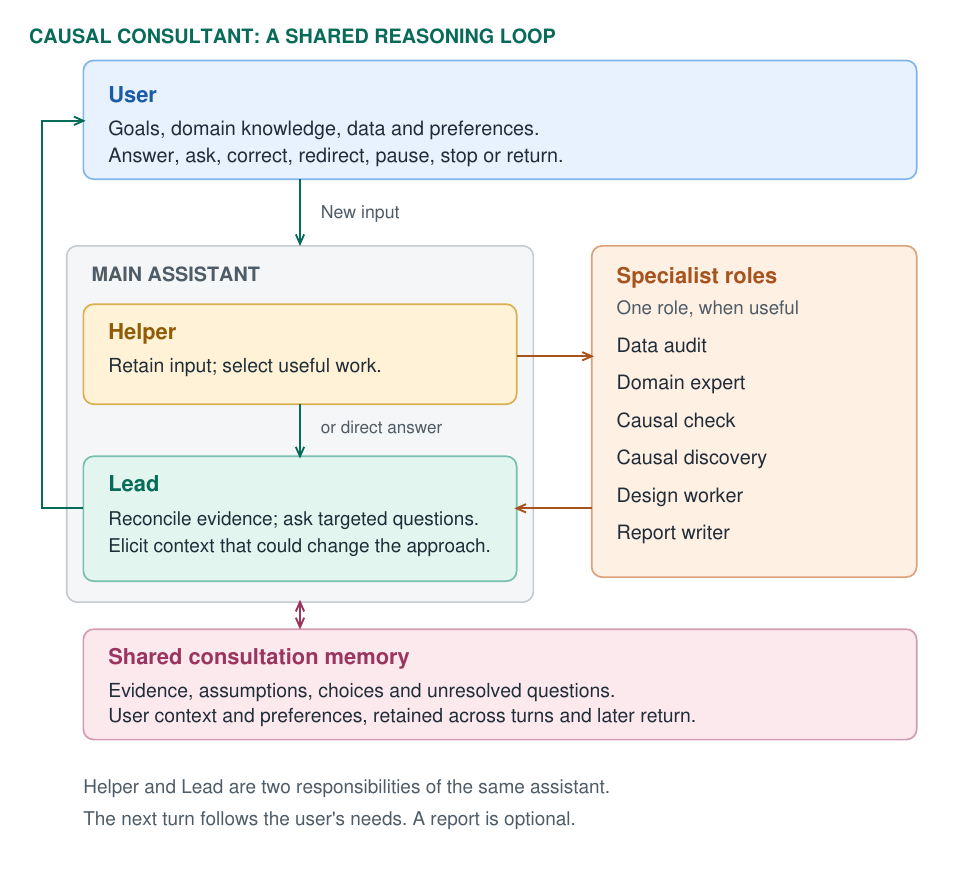}
\caption{The shared reasoning loop in the \emph{causal-consultant} skill (v8.0.7). Helper and Lead are responsibilities of the main assistant; a specialist contributes when useful. The assistant selects work, reconciles its consequences, and returns control to the user. Shared scientific memory connects successive rounds and later continuation. Arrows show responsibility and information flow, rather than a requirement for separate processes. The loop can produce an explanation, a revised question, an analysis, or an optional report \citep{zhu2026causalconsultant807}.}
\label{fig:causal-consultant-loop}
\end{figure}

Figure~\ref{fig:causal-consultant-loop} shows the organizing loop. \emph{Helper} and \emph{Lead} are two responsibilities of the same main assistant. Helper retains the incoming message, interprets its consequential facts, corrections, questions, and choices, and selects a useful next contribution. A direct explanation can proceed to Lead; a task needing specialized assessment can invoke one focused role. Lead's responsibility is to reconcile the returned assessment with the original objective and current evidence, assess which proposed changes are justified, and explain their consequences to the user. Specialist work can be performed within the main assistant or delegated when the host supports it. The design permits at most one substantive specialist review before returning control in a conversational round.

The user participates throughout this loop. They can supply a partial answer, request an explanation, correct a premise, change direction, pause, or return later. The consultation policy directs the assistant to ask for obtainable study facts and preferences, explain why they matter, and keep unresolved questions explicit when an answer is unavailable. It also distinguishes domain expertise, statistical expertise, and communication preferences so that explanations can be adapted without changing scientific standards. Concrete analysis and report proposals make their target, evidence, approach, and limitations available for the user's decision. There is no required sequence ending in estimation or a report: a useful round may end with a clarified question, a qualified recommendation, or a justified decision to stop.

\subsection{Engineering a Skill}

The preceding example makes skill design an engineering problem as well as a statistical one. A first version can focus on one recurring statistical decision, the study facts it requires, and contrasting examples of how different answers should change the work. Choices about context, persistent memory, specialist roles, and executable tools should follow from what that interaction requires and what the host supports. The \emph{causal-consultant} skill illustrates several choices that arise as a skill becomes more elaborate.

Context capacity is one such constraint. Some current models support windows of a million tokens or more \citep{google2026longcontext}, but this capacity does not guarantee equally reliable use of all the information supplied. Controlled evaluations show that performance can deteriorate as inputs grow, with effects that depend on the model, task, and distracting material \citep{hong2025context}. In a consultation, an accumulating history can also mix superseded assumptions, corrected facts, and repeated tool output. The author must therefore decide what belongs in the working context, what to retrieve on demand, and when to compact the history into a summary. For example, Anthropic's threshold-compaction API defaults to a trigger of 150,000 input tokens when enabled \citep{anthropic2026thresholdcompaction}. This is an operating policy, rather than a universal boundary for reliable reasoning. A suitable context budget and summarization policy need evaluation on the intended scientific tasks.

Memory design follows from this constraint. To support continuity, the \emph{causal-consultant} skill keeps an append-only history, an updated account of the study, and a concise summary for continuing the consultation. The engineering choice is to retain detailed evidence outside the working context while making relevant records retrievable. The statistician must specify which distinctions survive selection and summarization: observed facts, assumptions, preferences, unresolved questions, and conditions attached to results. For example, repeated district labels remain an observed feature of a dataset after the user explains that the labels were reused, but the interpretation that they identify stable units must change. A resumed consultation should recover both the correction and its consequence for an analysis requiring a stable panel. Saving every message alone does not ensure this; the design must connect storage, current interpretation, and retrieval.

Specialist roles introduce a related coordination problem. The author decides when a focused data audit or causal assessment is useful, what evidence it receives, and who incorporates its conclusions. Separate contexts can contain a specialist's detailed work without filling the main conversation, but the handoff must preserve consequential findings, assumptions, and unresolved alternatives. Multiple roles also create opportunities for duplicated work or incompatible interpretations. Their value depends on the division of responsibilities and the information passed between them.

Executable components likewise require choices about what software should enforce and what remains a matter of judgment. The memory utility preserves earlier versions and checks the consistency of stored records, while the agent must determine how new information changes the scientific interpretation. A successfully saved record can therefore contain an incorrect scientific interpretation. Computational tools need clear inputs, understandable errors, and outputs that expose the information needed for the next decision. A diagnostic summary might preserve a convergence warning and a reference to the complete fitted object, while keeping a lengthy numerical output outside the conversation. The statistician must determine which details are scientifically indispensable before the coding agent implements this interface.

The host also controls aspects of instruction loading, delegated contexts, compaction, and execution permissions \citep{anthropic2026}. A portable skill cannot assume that every host exposes the same controls. Its author should check, for example, whether a correction remains available after resumption, reaches the relevant specialist, and changes the appropriate recommendation. The statistician's engineering responsibility is to specify what information and behavior the scientific task requires, understand what the chosen host can support, and examine whether the two agree. Section~\ref{sec:next-generation} develops the broader design and evaluation choices when a custom harness is needed.

\section{Data Analysis, Study Design, and Statistical Interaction}
\label{sec:research-agent-harnesses}

Current agents provide several ways to connect language models, analytical tools, and human judgment. We reviewed independent software systems and applications with their own user interfaces for data analysis and study design, using primary evidence inspected through October 22, 2026. Appendix~\ref{app:analysis-systems} identifies the selected data analysis systems, their developers or authors, operating forms, and documented opportunities for human participation. Appendix~\ref{app:study-design-systems} separately summarizes selected applications that support study preparation before new data are collected. Both inventories include applications built on existing agent platforms and distinguish software availability from paper-described prototypes.

The reviewed systems span task delegation, conversational steering, and structured participation; Appendix~\ref{app:interaction-levels} reports their highest documented levels. It separately identifies \emph{human interaction guided by statistical principles}: interaction designed to examine assumptions, assess methodological appropriateness, or address risks of misleading analysis. Statistical principles guide what information the agent seeks, what the user is asked to consider, and how their contribution informs an analytical or study-design decision. We found qualifying documentation for seven of the 60 selected analysis systems. Optional modes count, and an unmarked system means that qualifying evidence was not established. These categories distinguish opportunities to participate and the purpose of participation; neither establishes routine use or scientific effectiveness.

\subsection{Data Analysis Agents and Human Participation}
\label{subsec:research-agent-automation}
\label{subsec:research-agent-interaction}

Automation remains a substantial objective. LAMBDA and TaskWeaver connect natural-language requests to executable analysis, while R\&D-Agent extends automation across model development and experimentation \citep{sun2025lambda,proposalTaskWeaver,proposalRDAgent}. Recent work continues this direction: Data Agents targets data tasks with minimal human intervention, with user confirmation as a fallback, and RADAR emphasizes typed operations, execution checks, and auditable results \citep{proposalDataAgent,proposalRADAR}. Automation and participation often coexist within the same application. Data Interpreter can pause for human editing and revise subsequent tasks, while AutoML-Agent offers optional review of requirements and proposed plans \citep{hong2024,inheritedNstatisticsautomlagent}. Their internal criticism or repair loops are not, by themselves, exchanges with a human, and successful execution does not establish that the analysis answers the intended scientific question.

Other systems organize analysis around shared work that people can inspect and revise. Deepnote Agent edits and executes notebook blocks, exposes changes, and continues from the revised notebook; Hex combines conversational analysis with notebooks and applications \citep{proposalDeepnote,proposalHex}. Data Formulator combines interface actions and natural language to support visualization and alternative analytical paths \citep{proposalDataFormulator}. TexeraAgent lets users revise executable dataflows or attach feedback to a past action and continue from that version, while ProCARE lets users revise a generated study plan before executing analyses of existing observational health data \citep{proposalTexeraAgent,proposalProCARE}. These interfaces provide opportunities to contribute substantive knowledge throughout the work.

Some interactions address methodological appropriateness directly. A Posit Assistant cleaning demonstration lets the user retain extreme flight delays as genuine observations \citep{proposalPositDemo}. ChatDA's default instructions call for discussing confounder selection with the user because it affects causal conclusions \citep{proposalChatDAPolicy}. CellVoyager reports expert feedback requesting donor-level pseudobulk analysis in place of a cell-level comparison, changing the unit of analysis \citep{inheritedC121}. MUSE surfaces warnings about questionable data treatment and supports user-selected repairs; its illustrated workflow revisits blanket removal of records with missing values \citep{chen2026muse}. Clarifying a desired threshold or business definition can also change an analysis, but does not by itself examine the statistical appropriateness of that choice.

Statistical expertise can also enter through the calculations and definitions that an application makes available. Posit's commons lets practitioners supply established calculations and supporting context, with a Shiny chat interface for using the configured analysis agent \citep{proposalCommons}. DSH Data Agent provides a database workbench and conversational analysis on DeepSeek Harness, with human review of business meanings and metric definitions \citep{proposalDSHDataAgent}. These applications illustrate how domain knowledge can inform analysis through reusable calculations or explicit definitions. They can build on existing runtimes while providing their own operating interfaces.

Before data collection, OrchestrXR describes human review and selective revision of experimental factors while preserving stable baselines, with revised specifications carried into an executable XR experiment \citep{proposalOrchestrXR}. This interaction is intended to preserve experimental control; the validity of resulting studies remains unestablished. TrialAtlas assesses trial designs and proposes improvements through a predominantly autonomous workflow \citep{proposalTrialAtlas}. Study preparation likewise requires distinguishing statistical computation, human control, and scientific justification.

\subsection{Designing and Evaluating Statistically Consequential Interaction}
\label{subsec:epistemic-interaction}

Existing applications provide concrete mechanisms for human participation, including interaction guided by statistical principles. The next question is whether that interaction supports a justified choice of target, assumption, method, uncertainty statement, or interpretation. The S category includes methodological review and user corrections as well as agent-initiated questions. Proactive inquiry is a particular design opportunity: the agent recognizes a consequential information gap, asks a question the user can answer, and connects the response to a statistical decision. An ordinary correction or preference can, however, be statistically consequential when it changes the target population or the relative costs of prediction errors. Statistical principles guide the interaction when the system helps examine the implications of that change.

\begin{figure}[!t]
\centering
\includegraphics[width=\textwidth]{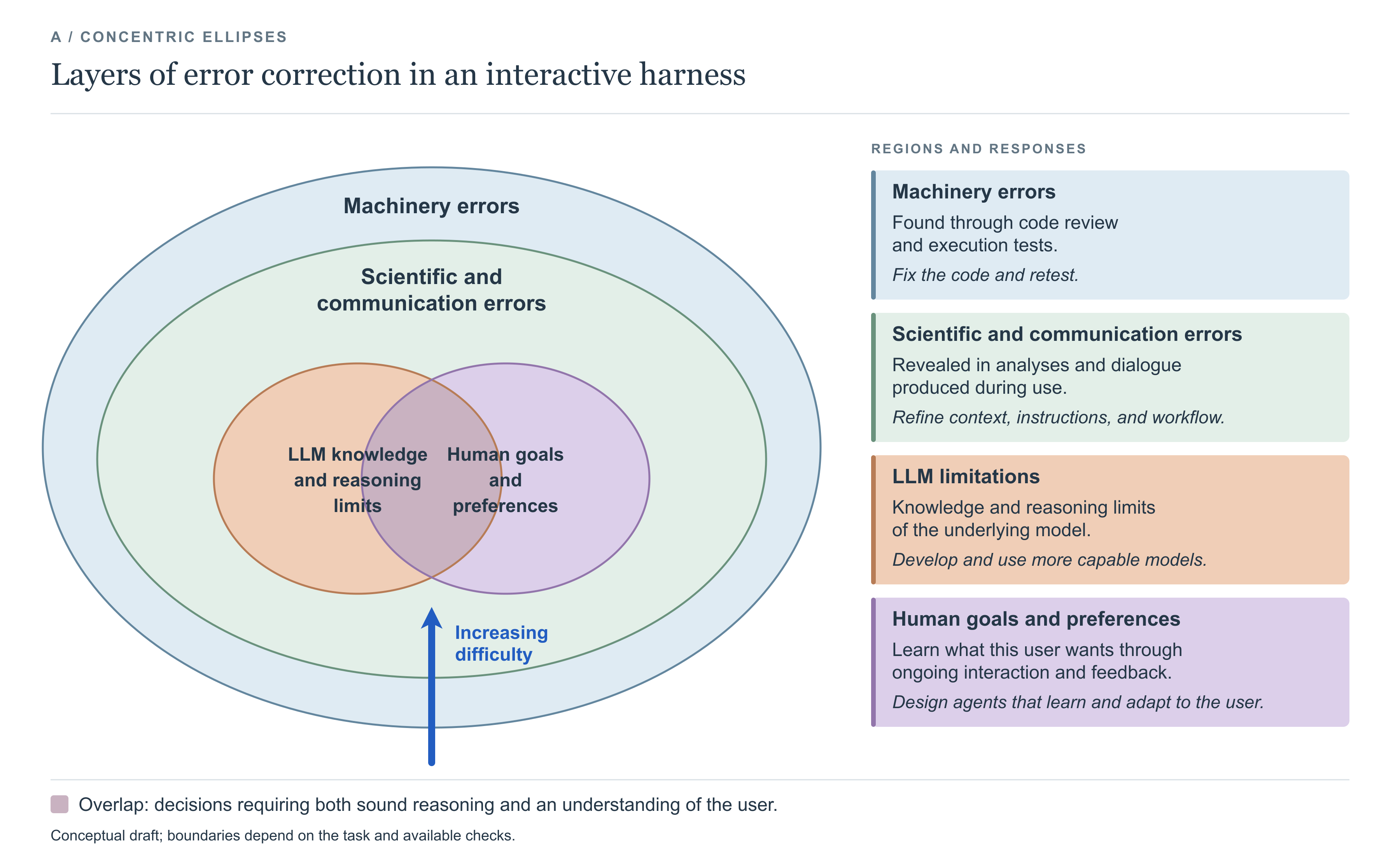}
\caption{A conceptual view of error correction in an interactive harness for statistical work. Explicit execution and numerical checks can make some failures easier to diagnose and repair. Scientific and communication failures require examining analyses and dialogue, while the overlapping inner regions emphasize decisions that depend on both model reasoning and an understanding of the user. The ellipses organize these challenges conceptually; they do not denote nested sets of errors or a universal ordering of difficulty. Human goals and preferences are conditions the system must understand, not errors to eliminate.}
\label{fig:harness-error-layers}
\end{figure}

For example, a request to compare outcomes between two programs can refer to a descriptive difference among current participants, prediction for future participants, or the effect of choosing one program over another. Clarifying that purpose changes what must be learned about enrollment, measurement, and follow-up, and what the data can support. An investigator may know that one program preferentially admitted more severe cases, or that outcomes were recorded only for participants who completed treatment. A statistical agent should connect these facts to selection, confounding, and the population represented by the observed outcomes. It should explain the resulting limitations or alternatives in the investigator's terms. The user supplies knowledge of the study; the system must still reason about its statistical implications, and an answer does not prove an identifying assumption.

Evaluation should examine both the usual workflow and optional interactive modes. Where feasible, it should compare whether consequential user information leads to an appropriate change in the work, relative to the same system without that information, while recording the effort required from the user. Relevant evidence connects a consequential information gap to an answerable question or correction, then to a justified change in the analytical work. A satisfactory response may instead preserve uncertainty, narrow the claim, or explain why the requested conclusion is unsupported. When the task is already well specified, further questioning may add effort without improving it. Thus neither the number of conversational turns nor the availability of approval, editing, or memory establishes the quality of the inquiry. Instructions express intended behavior, executable mechanisms make it possible, and evaluations must examine whether it occurs appropriately for the intended users and tasks.

Existing evaluation work supplies useful starting points. Ambig-DS studies clarification of ambiguous targets and objectives, and ConDABench evaluates conversational analysis with a simulated user \citep{proposalAmbigDS,proposalConDABench}. These settings make selected aspects of clarification measurable, while leaving open how systems handle evolving goals, incomplete knowledge, and misunderstandings in real consultation. Recent studies of methodological failures in agentic data science and of professionals using a general-purpose coding agent for data analysis further motivate assessing analytical validity and user understanding alongside task completion \citep{proposalSilentFailures,proposalExpertiseTrust}. They do not show that adding dialogue alone resolves these problems.

Figure~\ref{fig:harness-error-layers} organizes this evaluation problem by how failures can be diagnosed and corrected. Machinery errors, such as failed tool calls or incorrect state updates, and some numerical errors admit explicit checks against expected behavior or reference results. Once reproduced, they often permit a relatively direct cycle of correction and retesting. Scientific and communication errors require examining the analysis together with the dialogue: an unsupported generalization, a misunderstood answer, or a misleading explanation can persist even when the code executes correctly. Such failures may call for revisions to context, instructions, or the workflow, followed by evaluation of whether the revised system handles the underlying situation appropriately.

The harder problems arise where the model's knowledge and reasoning limits meet an incomplete understanding of the user's goals and preferences. Human preferences are not themselves errors; the system must learn what matters to the user while maintaining scientific justification. The appropriate next step depends jointly on the particular LLM and the evolving exchange and cannot always be specified in advance. More detailed instructions or a more capable model may help, but their adequacy must be evaluated in the intended conversations \citep{proposalConDABench}. Testing therefore needs to examine how the system adapts as users clarify or revise their goals, including when to stop asking or acknowledge a limitation. This makes the quality of the evolving interaction itself an object of statistical design and evaluation. The next section considers how to assemble and revise the components needed to support it.

\FloatBarrier

\section{A Second Entry Point:\texorpdfstring{\\}{ }Assembling a Custom Interactive Harness}
\label{sec:next-generation}

An interactive skill adapts an existing harness to express a statistical idea. The second entry point involves assembling a custom interactive harness when the host's available configuration and extensions cannot adequately support the intended behavior, scientific record, interface, or workflow. Much of the infrastructure can still be reused. We use \emph{runtime} for the execution machinery within the harness, including the loop that coordinates LLM calls and tool use. A builder can embed such a runtime and use a coding agent to implement the application components around it \citep{bonamy2026,openai2026codexdocs}.

\subsection{Model Forge: Interactive Model Development}
\label{subsec:model-forge}

The author's Model Forge (Codex edition, version 0.9.1) provides an implemented example of a domain-specific interactive harness for collaborative model and algorithm development \citep{rqzhu2026modelforge091}. It is a local application with a browser interface that uses the Codex runtime to support a Research Lead, Theorist, Data Analyst, and Outside Reviewer. Its aim is to develop a model for a stated problem through successive investigations and researcher decisions, rather than automatically generate a paper. A project can begin with a methodological question, a proposed formulation, or available data. The interface organizes work around problem formulation and literature exploration, method proposals, theoretical analysis, numerical evaluation, and a technical report. The researcher selects which candidate method and subsequent investigations to pursue; theory and numerical evaluation can inform each other in either order.

The instructions direct the team to seek input when missing study facts or researcher choices affect the formulation or justification of a method. As an illustration of the intended interaction, suppose a researcher describes the treatment, disease severity, and outcome variables collected for patients and asks how to estimate a treatment effect. The agent should recognize that the description omits when severity was measured relative to treatment initiation and outcome assessment. It could ask whether the value describes the patient before treatment or a later state that treatment may have changed. A value describing the patient after treatment would prompt reconsideration of a formulation that treated it as a baseline adjustment variable. The team should revisit candidate formulations and identification questions for the causal effect of interest, retaining alternatives when timing or causal relations remain unknown \citep{hernanrobins2020}. The Lead can handle routine decisions within the agreed scope, while the researcher supplies information needed to examine consequential assumptions.

The workbench makes formulations, assumptions, mathematical arguments, implementations, and numerical findings available for examination. It preserves the sources and versions underlying each investigation, including the model and theory to which numerical evidence refers. The technical report brings these results together with their limitations; an interim report can expose missing support and help identify further work. Report writing does not supply a missing proof or numerical result. Structural and numerical checks examine specified properties of the records, while scientific assessment remains dependent on model reasoning and researcher review. The example illustrates an implemented design; whether the team consistently elicits useful input and produces scientifically adequate work remains a question for evaluation with intended researchers and tasks.

\subsection{Infrastructure and Engineering Practices}
\label{subsec:builder-start}

Existing infrastructure offers several levels of control. The Codex software development kit (SDK) allows a statistical application to start and continue agent conversations. Codex app-server supports custom interfaces through which users can follow ongoing work, review requests for approval, and examine the conversation history \citep{openai2026codexdocs}. Agent frameworks offer another route. LangChain provides reusable model and tool integrations and configurable agent behavior, with its agents built on LangGraph \citep{langchain2026overview}. LangGraph exposes finer control over workflows that combine predetermined computational steps with model decisions, preserve state, and pause for human input \citep{langchain2026langgraph}. The choice depends on how much of the agent's execution behavior can be reused and how much the application needs to organize explicitly.

Open-source harnesses and platform extensions offer further scope for customization. DeepSeek Harness follows an ``everything is a plugin'' architecture built on Cordis, making the agent loop, model adapters, tools, and session records replaceable components \citep{deepseek2026harness}. Claude Mods lets developers alter prompts and tool calls and add interactive views within Claude Code; Anthropic's open-source playground provides sample implementations \citep{anthropic2026}. These mechanisms could support statistical inquiry through changes to both agent behavior and the interface through which users examine and correct its work.

Successful development also depends on how problems are framed, implementation is organized, and completion is judged. Community skill collections express complementary ways of thinking about these choices. Addy Osmani's \emph{Agent Skills} organizes development around specifications, incremental implementation, testing, review, and delivery \citep{osmani2026agentskills}. Matt Pocock's \emph{Skills for Real Engineers} emphasizes clarifying intent and resolving design questions before translating them into specifications and manageable implementation tasks \citep{pocock2026skills}. Lauren Tan's \emph{pstack} emphasizes investigation, checking actual behavior, and requiring evidence before declaring work complete, alongside coordination across agents \citep{tan2026pstack}. Taken together, these approaches suggest that plans should remain open to revision, implementation should proceed in small steps that can be checked, and completion should depend on evidence. For statisticians, this means connecting development to a scientific purpose, stating the assumptions the application relies on, and evaluating whether its behavior serves that purpose.

\begin{figure}[!htb]
\centering
\includegraphics[width=\textwidth]{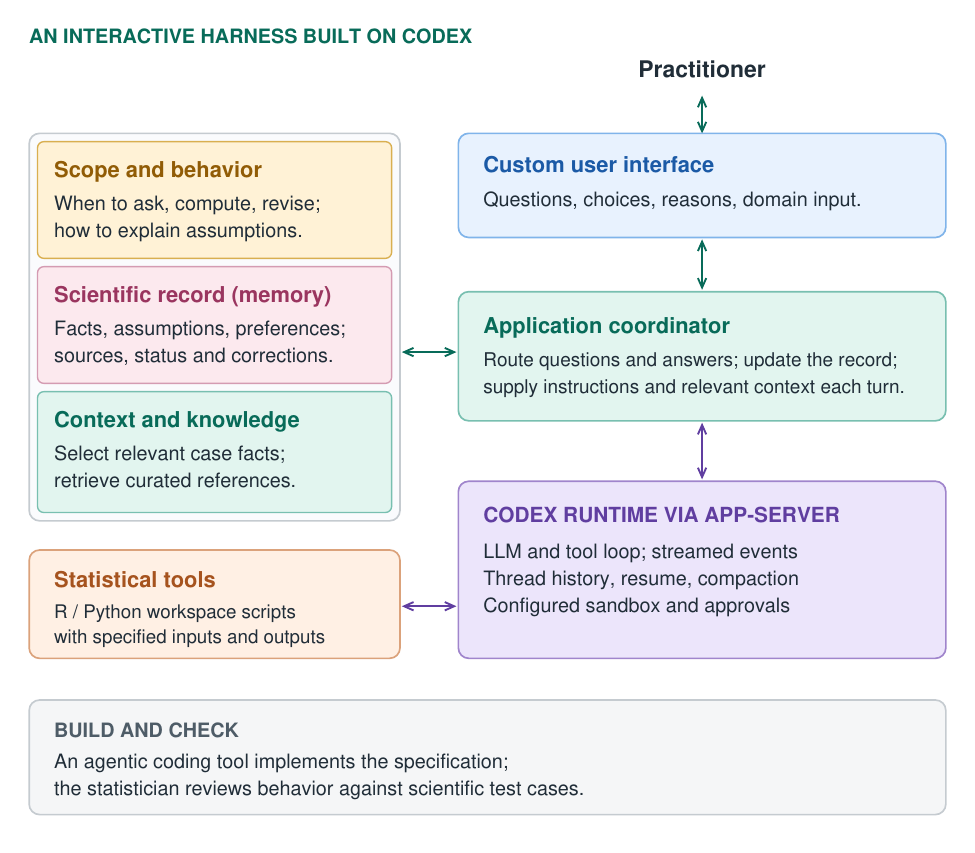}
\caption{An illustrative architecture for a custom interactive harness built on Codex combines application components with a reused runtime. The LLM itself is not shown. Colors distinguish behavior, scientific memory, context, interface, and tools; purple denotes the runtime and its connections. Arrows indicate exchanges of instructions, context, questions, answers, and results. Boxes describe logical responsibilities that can be reused or combined. The bottom strip concerns development and evaluation. The diagram summarizes reusable responsibilities. Codex capabilities follow its documentation and open-source implementation \citep{openai2026codexdocs,openai2026codexsource}.}
\label{fig:codex-statistical-harness}
\end{figure}

Figure~\ref{fig:codex-statistical-harness} illustrates one possible arrangement using the open-source Codex runtime \citep{openai2026codexdocs,openai2026codexsource}. The reused runtime and the application components together form an interactive harness for statistical work. Application code connects the inquiry instructions, scientific record, relevant knowledge, computational tools, and interface: it presents questions, records answers and corrections, selects context, and returns results. The following considerations concern the statistical purpose of these components and apply across infrastructure choices.

\subsection{Design Considerations}
\label{subsec:builder-considerations}

A custom harness needs a defined user, scientific task, and intended output, such as a study-design brief or a comparison of candidate models. The user's expertise shapes the explanation and assistance needed, while the task determines which decisions the agent can make and which require human judgment. Examples of an ordinary request, a missing fact, and a correction make the intended interaction concrete: what the user contributes and how the analysis, study plan, or model changes. Available data, computational resources, and the effort expected of the user also help determine a realistic scope. These choices shape the interface through which users examine and correct the evolving analysis, study plan, or model. A conversation alongside a visible study or model brief can show what is known, what remains open, and why a question matters. The interface should match the user's expertise: a practitioner may contribute study knowledge through concrete questions, while a methodological researcher may wish to examine formulations, challenge assumptions, and choose among research directions. Pausing and resuming with user input is supported by existing infrastructure \citep{langchain2026langgraph,openai2026codexdocs}; the statistical design determines when to ask and how an answer changes the record and subsequent work.

These interactions depend on what the system retains and brings into each decision. \emph{Context} is the information available to the model for its next action; \emph{memory} is information preserved for later use. Selection, retrieval, and compaction determine whether relevant facts remain available. Forgetting that treatment was assigned by clinic, for example, can change the interpretation of uncertainty. A scientific record should distinguish reported facts, assumptions, preferences, results, and unresolved questions, retaining their sources and subsequent corrections. Persistent memory can support a coherent account of the study across conversations, provided that corrections are connected to the assumptions and conclusions they affect \citep{langchain2026langgraph}. In methodological development, the record must also preserve which formulation, assumptions, and implementation each theoretical or numerical result concerns. Revising a statistical model should not silently transfer support established for an earlier version. This scientific context also determines which computational tools are appropriate. Existing statistical packages, scripts, and numerical libraries can supply much of the computation, with clear inputs, inspectable outputs, and an account of when each operation is appropriate. For example, DoWhy separates causal modeling, identification, estimation, and validation, providing computational stages that an agent can connect to questions about their assumptions \citep{sharma2021causalassumptions}. Retained code, inputs, and results should remain linked to their supporting assumptions and user contributions so that revisions can be traced to affected conclusions. The record should also distinguish revisions prompted by newly supplied study information from choices made after inspecting analytical results. Repeatedly changing models, outcomes, or hypotheses in response to the same data can introduce selection effects and hidden multiplicity. The harness should make such exploration visible and connect subsequent claims to an appropriate validation strategy; recording the sequence alone does not validate the final inference \citep{sosa2026}.

The choice and combination of AI models also affect response speed and the reliability of the decisions assigned to the agent. TypeSafe's Jev, for example, is designed for fast, constrained decisions rather than text generation \citep{almeida2026jev}; such models could support bounded judgments alongside LLMs used for broader reasoning and explanation. Speed and structured outputs alone do not establish scientific reliability. Routing between LLMs already offers a way to balance cost and response quality \citep{ong2025routellm}. If training or adapting LLMs becomes substantially easier, domain-specific models could be trained to be more sensitive to a field's concepts, assumptions, and common sources of error. A harness could call such models for relevant tasks alongside models used for broader reasoning and communication, with each assessed for its intended role. These choices also raise questions about decision stability: irrelevant changes in wording should not alter a recommendation, whereas new study information may warrant revision. The harness should make its scientific work inspectable so that such differences can be examined. Such traceability links the design of the tools and scientific record to the testing and revision discussed next.

\subsection{Testing, Validation, and Revision}
\label{subsec:builder-validation}

The evaluation criteria in Section~\ref{subsec:epistemic-interaction} can be translated into tests of tools, retained information, and conversational behavior. Numerical tools can be checked against reference results; contrasting conversations can test whether corrections reach subsequent analyses and whether missing information prompts an appropriate response. Because users may supply incomplete answers or revise their goals, these tests should allow for more than one scientifically justified response.

Testing should include missing information, conflicting answers, unsupported requests, and changes in the scientific question, examining whether the agent clarifies, revises, offers alternatives, or defers judgment appropriately. Such cases cannot exhaust unfamiliar exchanges, where the LLM must still recognize consequential information and reason about it.

Validation and revision should proceed together. Contrasting cases can test whether the same numerical data lead to different analyses when the observation process differs. In the district example from Section~4.3, one response could confirm that identifiers denote stable geographical units, while another could explain that the codes were reassigned. Evaluation would examine whether the agent uses that distinction when considering a panel analysis, preserves unresolved requirements, and carries the correction into later work. For the timing example in Section~\ref{subsec:model-forge}, evaluation could examine whether a corrected measurement history changes the candidate formulations and identification questions, and whether the correction remains available when the investigation resumes. It should also examine whether the system requests consequential input without repeatedly returning routine decisions to the researcher. Cases needing no further questions can reveal unnecessary interruption. Repeated trials help assess variation, and simulation can examine uncertainty statements when the generating process is known. These checks need to be complemented by review of conversations and evaluation with intended users, since scripted cases cannot exhaust the ways people explain, misunderstand, or revise a problem. Observed failures can motivate changes to instructions, tools, memory, interface, or model choice, with earlier cases rerun to check whether the revision helps without introducing new errors. How best to divide responsibility between explicit design and model judgment remains an empirical question for statistical applications.

\section{Looking Ahead: Shaping Agentic AI\texorpdfstring{\\}{ }Through Statistical Thinking}
\label{sec:what-to-build}

Causal Consultant and Model Forge illustrate what statisticians can build for research, but the same opportunity extends to people building tools for themselves. Suppose someone wants to fine-tune a language model on past conversations so that it can draft replies in their style \citep{hauser2026whatsapp}. Before training, an agent might ask, ``Do you want help replying to people you already know, or should this also work when you write to someone new?'' For the first use, the agent could reserve later exchanges with familiar correspondents for evaluation. For the second, it could reserve entire conversations with people whose messages the model never saw during training. The user can say which task matters and judge whether a suggested reply is one they would send. The statistician's contribution is to make that answer determine how the model is evaluated. A harder question follows: what should the agent ask at the outset, and what can it learn only after the person sees a draft and reconsiders what they want?

One useful perspective treats the human as a partially observed, evolving environment. Let $X_t$ denote the context available at stage $t$, including the interaction history, and let $Z_t$ represent the much larger, potentially infinite-dimensional body of information about the user and task that remains partly unobserved. The agent acts using $X_t$ and can elicit information about previously unobserved aspects of $Z_t$ through questions or other actions. User responses can update $X_t$, guide subsequent choices, and provide noisy, incomplete evidence about whether previous actions helped. The agent's output may also change the user's understanding, preferences, and goals, so aspects of $Z_t$ and the way the user responds can themselves evolve. Facts about how a completed study was conducted remain fixed, although what the agent knows about them can change. Questioning, acting, and responding thus evolve together, while outcomes can depend on information that is never elicited. Existing literature examines aspects of this process through different modeling frameworks and assumptions. Experiential preference elicitation relates the cost and noise of user responses to their context \citep{perrault2019elicitation}; active-measurement models in reinforcement learning allow measurement itself to affect future states \citep{gao2026activemeasuring}. Budgeted contextual bandits study information acquisition under a limited budget \citep{gan2024budgetedreveal}, while approximate information states support planning using representations of observed history \citep{subramanian2022information}. These approaches offer useful starting points, with their conclusions depending on assumptions about what can be observed, how states and responses evolve, and how outcomes are evaluated.

Quantifying the overall behavioral process in consultation and collaborative methodological research remains difficult. A conversation may reveal missing study information while changing the scientific question the user wishes to pursue. The agent must judge what to ask and when further inquiry is unlikely to improve the outcome enough to justify the user's effort, even as preferences and understanding change. Feedback may be selective, delayed, or absent, and immediate satisfaction need not indicate scientific benefit. Learning from an individual user may involve only a short interaction history, even when the agent draws on extensive training data. This can make adaptation to the user a small-sample problem: uncertainty about the user's context and the consequences of alternative actions should guide what the agent asks and does next. Adaptive questioning, dependence between responses, and differences across users also complicate evaluation and uncertainty quantification. For example, independent evaluations of a fixed harness on tasks sampled from a specified population could record whether each task produces a prespecified methodological error. Confidence sequences could quantify uncertainty about that error rate as evaluations accumulate \citep{howard2021confidencesequences}; extending such assessment to evolving dialogues with selective feedback remains a research challenge.

Progress will require joint efforts from statistics, linguistics, education, psychology, human-computer interaction, and machine learning, including reinforcement learning. Research on conversational grounding and adaptive tutoring can inform how agents communicate and support learning \citep{clark1991grounding,alam2023hints}. Statistical principles can help turn these insights into informative measurements, reliable evaluations, and decisions under uncertainty. By helping build these agents, statisticians can make a good statistical question available to someone who would never have known to ask it, at the moment when the answer can still change the analysis.

\section*{Acknowledgements}

The author used OpenAI Codex to assist with writing and editing this manuscript.

\PaperMainCitationsfalse
\bibliographystyle{plainnat}
\PaperPrintMainRefstrue
\begin{paperreferences}{References}
\bibliography{references}
\end{paperreferences}

\clearpage
\section*{Appendix}
\appendix
\section{Selected Applications for Data Analysis}
\label{app:analysis-systems}
\setcounter{table}{0}
\renewcommand{\thetable}{A.\arabic{table}}
\renewcommand{\theHtable}{A.\arabic{table}}
\setcounter{figure}{0}
\renewcommand{\thefigure}{A.\arabic{figure}}
\renewcommand{\theHfigure}{A.\arabic{figure}}

The inventory was initially compiled using GPT-6 Astra with the Ultra thinking setting over multiple rounds of extensive searches. Additional AI-assisted searches and source checks drew on primary papers, repositories, and official documentation. The review includes independent software applications or systems with their own user interface, including applications built on another agent platform, and excludes skill-only packages. These checks did not include runtime testing or independent reproduction of reported scientific evaluations. The author does not guarantee the accuracy of all information reported in this appendix. Readers are welcome to contact the author with suggestions for additions or corrections.

Table~\ref{tab:analysis-systems} presents 60 selected applications and research systems identified using primary evidence inspected through October 22, 2026. It covers data preparation, visualization, statistical analysis, and machine-learning development, with selected domain applications. Command-line and notebook-launched research applications qualify. APIs or libraries without an application, generic coding hosts, pure literature search, and proof systems are excluded. Paper-described application prototypes are labeled when public software is unverified. This purposive inventory is not a census, ranking, or count of new releases.

Names, attribution, interfaces, and links were checked against primary papers, repositories, package metadata, and official documentation. Verified developers, paper authors, or explicitly identified repository maintainers provide the attribution; related product surfaces are grouped. Repository links identify accessible source, without asserting a license or successful execution. Hosted products may require accounts or payment. Rows distinguish public source, hosted products, and paper-described applications for which public software availability was not verified. Benchmarks and human studies inform Section~\ref{sec:research-agent-harnesses} but are excluded from this table.

\subsection{Levels of Human Interaction}
\label{app:interaction-levels}

Each system is assigned the highest level of human participation established by the reviewed sources. The levels distinguish forms of participation, rather than scientific quality or suitability: Level 1 can be appropriate for a well-specified task, and Level 3 need not be appropriate for every user. Optional modes and conditional checkpoints count, so a Level 3 classification does not imply that routine runs are highly interactive. Internal agent debate, automated error correction, and generic execution permissions do not count as substantive human participation. The classification describes documented capabilities of the reviewed versions, not the frequency with which people use them.

\begin{description}
\item[Level 1: Task delegation.] A person supplies a task, data, or configuration and receives results. Ongoing human participation in the analytical work is not established by the reviewed sources.
\item[Level 2: Conversational steering.] A person can ask follow-up questions, correct the request, or redirect and rerun the analysis through dialogue.
\item[Level 3: Structured workflow participation.] A person can review or revise analytical plans or intermediate stages, edit shared analytical artifacts through the operating application, or compare and pursue alternative analytical paths. The edited artifacts must enter subsequent work; these can include code, data, or reusable metric definitions. A chat interface alone does not establish this level.
\end{description}

\emph{Human interaction guided by statistical principles} refers to interaction designed to examine assumptions, assess methodological appropriateness, or address risks of misleading analysis. Statistical principles guide what information the agent seeks, what the user is asked to consider, and how their contribution informs an analytical or study-design decision. This form of interaction is coded separately. A superscript \textsuperscript{S} requires documentation of three connected elements: (i) a specific concern about statistical validity or appropriate data treatment or interpretation; (ii) a substantive human role in examining that concern, through contextual information, assessment of assumptions, or methodological review; and (iii) a route by which that contribution can lead to retaining, revising, restricting, or rejecting an analytical or study-design choice. Relevant concerns include dependence, selection, confounding, measurement, missingness, validation, and uncertainty. An explicit operating policy can establish the intended purpose when linked to an available interaction mechanism; an implemented or illustrated exchange provides more direct evidence. Adaptive questioning is not required.

Clarifying a target column, choosing a missingness threshold, completing calculator inputs, or resolving a business definition does not qualify by itself. Nor do generic approval, editable outputs, or automated statistical checks without a substantive human role. For example, asking which rows to remove differs from examining whether removing them would distort the analysis. The mark concerns the documented purpose and mechanism of the interaction, not whether it reliably improves statistical validity. An unmarked system means that qualifying evidence was not established, not that the capability is absent.

\subsection{Years and Classification Summary}
\label{app:interaction-years}

\begin{figure}[!htbp]
\centering
\includegraphics[width=\textwidth]{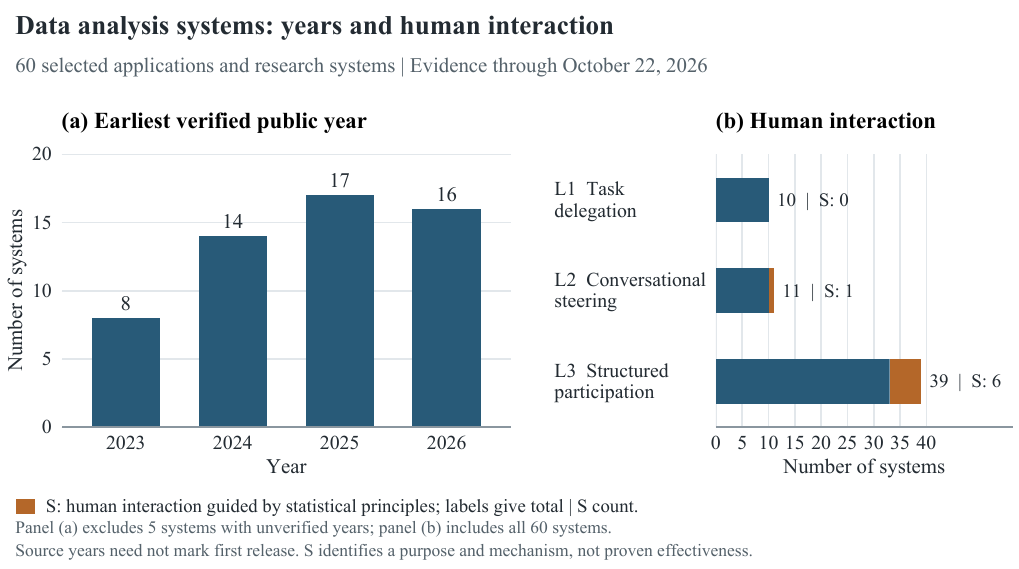}
\caption{The selected data analysis systems, using primary evidence inspected through October 22, 2026. Left: counts by the year of the earliest verified public evidence found; systems with unverified years are omitted. Right: documented human-interaction levels for all 60 systems; highlighted segments identify the seven systems with documented human interaction guided by statistical principles. Unhighlighted segments indicate that qualifying evidence was not established. The levels describe forms of participation; the S mark identifies a purpose and mechanism, not demonstrated scientific effectiveness.}
\label{fig:analysis-systems-summary}
\end{figure}

Figure~\ref{fig:analysis-systems-summary} summarizes verified years of public evidence and interaction levels for all 60 systems; systems with unverified years are omitted from the annual bars. The year is the earliest verified public evidence found for the listed analytical system or capability, using dated papers, official product announcements, releases, or dated software artifacts. A parent platform's launch, repository creation or historical commit date alone, current citation year, and review date are not substitutes. Paper dates can establish that a research system was described without establishing a software release. These years do not necessarily identify first availability. Related versions and renames count once. Interaction levels describe the reviewed capabilities, not necessarily the capabilities at the recorded year. The purposive sample and incomplete release histories do not support estimates of market growth or prevalence.

The inventory contains 10 systems at Level 1, 11 at Level 2, and 39 at Level 3. Human interaction guided by statistical principles is documented for seven systems: one at Level 2 and six at Level 3. These counts summarize the source-based classifications, not a comparison of analytical accuracy.

\FloatBarrier

\subsection{Categorized Inventory}
\label{app:categorized-inventory}

Table~\ref{tab:analysis-systems} groups systems by interaction level. Each row gives the operating form and a short description of documented participation. The participation description, year, and superscript \textsuperscript{S} link to supporting primary sources; the final column links to the product, repository, or paper. Bibliographic years in author citations can differ from the year of the earliest public evidence found. ``n.v.'' denotes an unverified year.

\begingroup
\footnotesize
\setlength{\tabcolsep}{3pt}
\renewcommand{\arraystretch}{1.12}
\begin{longtable}{@{}L{0.19\textwidth}L{0.25\textwidth}L{0.32\textwidth}L{0.06\textwidth}L{0.10\textwidth}@{}}
\caption{Selected data analysis systems, categorized by documented human interaction. \textsuperscript{S}: documented human interaction guided by statistical principles.}\label{tab:analysis-systems}\\
\toprule
System & Developer / authors (reference) & Form / documented participation & Year & Link \\
\midrule
\endfirsthead
\multicolumn{5}{l}{\textit{Table \thetable\ continued}}\\
\toprule
System & Developer / authors (reference) & Form / documented participation & Year & Link \\
\midrule
\endhead
\midrule
\multicolumn{5}{r}{\footnotesize Continued on next page}\\
\endfoot
\bottomrule
\endlastfoot
\multicolumn{5}{@{}l}{\textbf{Level 1: Task delegation (10 systems)}} \\*[4pt]
AIDE & WecoAI \citep{inheritedNstatisticsaide} & \textit{Runnable ML experiment application}.\newline \href{https://github.com/WecoAI/aideml/blob/main/README.md}{Specify goal and metric; inspect the autonomous search and resulting code.} & \href{https://pypi.org/project/aideml/0.1.0/}{2024} & \href{https://github.com/WecoAI/aideml}{Repository} \\[3pt]
AIRA\_GREEDY (AIRA-dojo application) & Meta / Facebook Research \citep{inheritedNstatisticsairadojo} & \textit{Command-line experiment runner; results visualization}.\newline \href{https://github.com/facebookresearch/aira-dojo/blob/main/README.md}{Configure a task and budget; autonomous experiments produce scored solutions.} & \href{https://arxiv.org/abs/2507.02554}{2025} & \href{https://github.com/facebookresearch/aira-dojo/blob/main/src/dojo/configs/_exp/mlebench/aira_greedy_o3.yaml}{Repository} \\[3pt]
AutoBA & \citet{inheritedC124} & \textit{Configuration-driven command-line application}.\newline \href{https://github.com/JoshuaChou2018/AutoBA/blob/main/README.md}{Specify files and goals; the application plans and executes the analysis.} & \href{https://github.com/JoshuaChou2018/AutoBA/blob/main/README.md}{2023} & \href{https://github.com/JoshuaChou2018/AutoBA}{Repository} \\[3pt]
AutoMind & \citet{proposalAutoMind} & \textit{Command-line ML research application; public source}.\newline \href{https://github.com/zjunlp/AutoMind/blob/1a1ba234ee2bd1e831c6757cd2ec01c2e4ccf52e/README.md}{Configure a task and budget for autonomous code search and refinement.} & \href{https://arxiv.org/abs/2506.10974v1}{2025} & \href{https://github.com/zjunlp/AutoMind}{Repository} \\[3pt]
Data-Copilot & \citet{inheritedNstatisticsdatacopilot} & \textit{Gradio web application and runnable backend}.\newline \href{https://github.com/ZJU-OmniAI/Data-Copilot/blob/main/README.md}{Submit an analytical request; inspect generated steps, tables and charts.} & \href{https://arxiv.org/abs/2306.07209}{2023} & \href{https://github.com/ZJU-OmniAI/Data-Copilot}{Repository} \\[3pt]
DS-Agent & \citet{inheritedNstatisticsdsagent} & \textit{Command-line research application}.\newline \href{https://github.com/guosyjlu/DS-Agent/blob/main/README.md}{Supply a task; automated case retrieval, experimentation and refinement follow.} & \href{https://arxiv.org/abs/2402.17453}{2024} & \href{https://github.com/guosyjlu/DS-Agent}{Repository} \\[3pt]
ML-Master (public v1 implementation) & \citet{inheritedNstatisticsmlmaster} & \textit{Command-line research application}.\newline \href{https://github.com/sjtu-sai-agents/ML-Master/blob/main/README.md}{Configure a task and budget; autonomous exploration returns candidate solutions.} & \href{https://github.com/sjtu-sai-agents/ML-Master/blob/main/README.md}{2025} & \href{https://github.com/sjtu-sai-agents/ML-Master}{Repository} \\[3pt]
MLE-STAR & \citet{proposalMLESTAR} & \textit{Paper-described ML application; current public software unverified}.\newline \href{https://arxiv.org/abs/2506.15692v3}{Supply a task and data for autonomous model search, component refinement, and ensembling.} & \href{https://research.google/blog/mle-star-a-state-of-the-art-machine-learning-engineering-agents/}{2025} & \href{https://arxiv.org/abs/2506.15692v3}{Paper} \\[3pt]
RADAR & \citet{proposalRADAR} & \textit{Research runtime; public software unverified}.\newline \href{https://arxiv.org/html/2609.32528v1}{Submit a task and data; automated runtime checks guide the agent's analysis and repairs.} & \href{https://arxiv.org/html/2609.32528v1}{2026} & \href{https://arxiv.org/abs/2609.32528}{Paper only} \\[3pt]
SELA & \citet{inheritedNstatisticssela} & \textit{Command-line experiment application}.\newline \href{https://github.com/FoundationAgents/MetaGPT/blob/main/metagpt/ext/sela/README.md}{Configure the ML task; autonomous tree search explores analytical pipelines.} & \href{https://arxiv.org/abs/2410.17238}{2024} & \href{https://github.com/FoundationAgents/MetaGPT/tree/main/metagpt/ext/sela}{Repository} \\[3pt]
\multicolumn{5}{@{}l}{\textbf{Level 2: Conversational steering (11 systems)}} \\*[4pt]
Alias Data Science mode & \citet{inheritedNstatisticsaliasdatascience} & \textit{CLI and full-stack application interface}.\newline \href{https://github.com/agentscope-ai/agentscope-samples/blob/main/alias/README.md}{Interrupt and redirect analytical work through a stateful application conversation.} & \href{https://github.com/agentscope-ai/agentscope-samples/blob/main/alias/README.md}{2025} & \href{https://github.com/agentscope-ai/agentscope-samples/tree/main/alias}{Repository} \\[3pt]
BambooAI & Palo Galko \citeyearpar{proposalBamboo} & \textit{Public analysis application; web UI}.\newline \href{https://github.com/pgalko/BambooAI/blob/11698d8e5c794f002e6db0f2ae9684a3456127a0/analyst/contract.md}{Continue analysis through follow-up questions, corrections, and explicit requests for user clarification.} & \href{https://github.com/pgalko/BambooAI/releases/tag/v0.0.7}{2023} & \href{https://github.com/pgalko/BambooAI}{Repository} \\[3pt]
Biomni & \citet{inheritedC077} & \textit{Python application and Gradio/web interface}.\newline \href{https://github.com/snap-stanford/Biomni/blob/main/biomni/agent/a1.py}{Continue biomedical analyses through a conversation that retains previous user and assistant turns.} & \href{https://cs.stanford.edu/people/jure/pubs/biomni-biorxiv25.pdf}{2025} & \href{https://github.com/snap-stanford/Biomni}{Repository} \\[3pt]
ChatDA / TableMage\href{https://github.com/YMa-lab/TableMage/blob/a1fa00787ea1beccb6a1f79b6e73ef1cae514bd6/tablemage/agents/_src/agents_src/prompt/single_agent_system_prompt.py}{\textsuperscript{S}} & \citet{proposalChatDA} & \textit{Python analysis application with its own Flask web UI}.\newline \href{https://github.com/YMa-lab/TableMage/blob/a1fa00787ea1beccb6a1f79b6e73ef1cae514bd6/tablemage/agents/_src/agents_src/prompt/single_agent_system_prompt.py}{Continue analysis over retained data; default instructions call for discussing confounder selection with the user because it affects causal conclusions.} & n.v. & \href{https://github.com/YMa-lab/TableMage}{Repository} \\[3pt]
commons & Posit \citep{proposalCommons} & \textit{Configurable analysis agent; Shiny chat UI}.\newline \href{https://posit-dev.github.io/commons/r/articles/commons.html}{Chat with a deployed data agent grounded in curated calculations and analytical context.} & \href{https://pypi.org/project/commons/}{2026} & \href{https://posit-dev.github.io/commons/}{Docs} \\[3pt]
CompBioAgent & BioInfoRx / Biogen / Stanford \citep{inheritedNbio10} & \textit{Web application}.\newline \href{https://github.com/interactivereport/CompBioAgent/blob/main/webapp/app/core/app_ai_assistant_chat_content.php}{Refine biological queries and rerun visualizations with changed user settings.} & \href{https://bioinforx.com/web/news.php?id=1000101}{2025} & \href{https://github.com/interactivereport/CompBioAgent}{Repository} \\[3pt]
Cortex Agents / CoWork & Snowflake \citeyearpar{proposalSnowflake} & \textit{Warehouse agents and hosted interface}.\newline \href{https://docs.snowflake.com/en/user-guide/snowflake-cortex/snowflake-cowork}{Ask follow-ups and continue investigations from saved analytical outputs.} & \href{https://docs.snowflake.com/en/release-notes/2025/other/2025-08-01-snowflake-intelligence}{2025} & \href{https://docs.snowflake.com/en/user-guide/snowflake-cortex/cortex-agents}{Docs} \\[3pt]
Data-Juicer Agent & \citet{inheritedNstatisticsdatajuiceragent} & \textit{Runnable conversational application}.\newline \href{https://github.com/agentscope-ai/agentscope-samples/blob/main/data_juicer_agent/main.py}{Iteratively request and adjust data processing through a persistent conversation.} & n.v. & \href{https://github.com/agentscope-ai/agentscope-samples/tree/main/data_juicer_agent}{Repository} \\[3pt]
Edison Analysis (formerly Finch) & Edison Scientific \citep{inheritedNbio17} & \textit{Hosted analytical workspace and API}.\newline \href{https://edisonscientific.com/news/introducing-edison-analysis}{Ask follow-up questions after notebook analysis; download auditable results.} & \href{https://edisonscientific.com/news/introducing-edison-analysis}{2025} & \href{https://edisonscientific.com/news/introducing-edison-analysis}{Product} \\[3pt]
SpatialAgent & Genentech-led team \citep{inheritedC128} & \textit{Runnable Python/notebook analysis agent}.\newline \href{https://tawaunlucas.com/publication/spatialagent/SpatialAgent.pdf}{Use co-pilot mode to refine task definitions and ask follow-up questions during analysis.} & \href{https://www.biorxiv.org/content/10.1101/2025.04.03.646459v1}{2025} & \href{https://github.com/Genentech/SpatialAgent}{Repository} \\[3pt]
TaskWeaver data-analysis application & Microsoft \citep{proposalTaskWeaver} & \textit{CLI and UI clients; archived}.\newline \href{https://github.com/microsoft/TaskWeaver/blob/main/README.md}{Steer multi-round analyses while conversation and execution state persist.} & \href{https://arxiv.org/abs/2311.17541}{2023} & \href{https://github.com/microsoft/TaskWeaver}{Repository} \\[3pt]
\multicolumn{5}{@{}l}{\textbf{Level 3: Structured workflow participation (39 systems)}} \\*[4pt]
AdaLens & \citet{liu2026adalens} & \textit{Research web application; public software unverified}.\newline \href{https://arxiv.org/html/2608.17834v1}{Modify analytical plans, focus or suppress topics, and pause or resume individual analysis threads.} & \href{https://arxiv.org/html/2608.17834v1}{2026} & \href{https://arxiv.org/abs/2608.17834}{Paper only} \\[3pt]
AgenMed\_CSS\href{https://github.com/boccxx/AgenMed_CSS/blob/ffe909d6db07d4b5c1a855315d6cfd47cdb4cf1c/frontend/protocol_generation.html}{\textsuperscript{S}} & \citet{proposalAgenticTSTROBE} & \textit{Planning and analysis web application; public source}.\newline \href{https://github.com/boccxx/AgenMed_CSS/blob/ffe909d6db07d4b5c1a855315d6cfd47cdb4cf1c/frontend/protocol_generation.html}{Review measurement validity and comparison-group suitability in guided forms; revise the protocol/SAP used for analysis. Demonstration uses existing data.} & \href{https://www.maxapress.com/article/id/6ac6e6b9fa6c585e89e1f595}{2026} & \href{https://github.com/boccxx/AgenMed_CSS}{Repository} \\[3pt]
AI Pipeline Studio & Business Science \citep{inheritedNstatisticspipelinestudio} & \textit{Streamlit web application}.\newline \href{https://github.com/business-science/ai-data-science-team/blob/master/apps/ai-pipeline-studio-app/README.md}{Edit and rerun pipeline steps; compare versions; clarify modeling targets.} & \href{https://github.com/business-science/ai-data-science-team/releases/tag/0.0.0.9017}{2025} & \href{https://github.com/business-science/ai-data-science-team/tree/master/apps/ai-pipeline-studio-app}{Repository} \\[3pt]
AutoKaggle & \citet{inheritedNstatisticsautokaggle} & \textit{Runnable staged ML application}.\newline \href{https://arxiv.org/html/2410.20424v3}{Optionally review and refine analytical plans at successive workflow stages.} & \href{https://arxiv.org/abs/2410.20424}{2024} & \href{https://github.com/multimodal-art-projection/AutoKaggle}{Repository} \\[3pt]
AutoML-Agent & \citet{inheritedNstatisticsautomlagent} & \textit{Notebook/Python-launched research application}.\newline \href{https://github.com/DeepAuto-AI/automl-agent/blob/9e0d842959a09d96c00768c14d73bd9df6d39023/agent_manager/__init__.py}{Optionally review interpreted requirements and proposed plans before allowing analysis to proceed.} & \href{https://arxiv.org/abs/2410.02958}{2024} & \href{https://github.com/DeepAuto-AI/automl-agent}{Repository} \\[3pt]
AutoSTAT v2 & Jiaye-s-Group \citeyearpar{proposalAutoSTAT} & \textit{Public statistical analysis application}.\newline \href{https://github.com/Jiaye-s-Group/AutoSTAT/blob/3343761d7c90458f4bc08ae2e05bcdd8fdc04759/frontend/utils/suggestion_state.py}{Review, revise, and confirm staged analytical suggestions before generating or rerunning code.} & n.v. & \href{https://github.com/Jiaye-s-Group/AutoSTAT}{Repository} \\[3pt]
BioMANIA & BATMEN Lab, Waterloo \citep{inheritedNbio08} & \textit{Web and CLI conversation}.\newline \href{https://github.com/batmen-lab/BioMANIA/blob/main/src/deploy/model.py}{Review proposed analysis operations, revise parameters, and return to previous conversational states.} & \href{https://github.com/batmen-lab/BioMANIA/blob/main/docs/version_history.md}{2023} & \href{https://github.com/batmen-lab/BioMANIA}{Repository} \\[3pt]
CARIS\href{https://arxiv.org/html/2604.12258v2}{\textsuperscript{S}} & \citet{proposalCARIS} & \textit{Paper-described NiceGUI web prototype; later React source public}.\newline \href{https://arxiv.org/html/2604.12258v2}{Review and revise research plans used downstream; reported expert revisions removed inappropriate analytical components.} & \href{https://arxiv.org/abs/2604.12258v2}{2026} & \href{https://arxiv.org/abs/2604.12258v2}{Paper} \\[3pt]
CASSIA & \citet{inheritedNbio07} & \textit{Web application and CLI; R/Python packages also available}.\newline \href{https://github.com/ElliotXie/CASSIA/blob/main/docs/CASSIA_Tutorial_Web/Cassia_Documentation_English/content/en/docs/python/annotation-boost.md}{Review annotations, choose uncertain clusters, and launch targeted refinement or comparisons of candidate identities.} & \href{https://github.com/ElliotXie/CASSIA/releases/tag/CASSIA}{2024} & \href{https://github.com/ElliotXie/CASSIA}{Repository} \\[3pt]
CellVoyager\href{https://media.springernature.com/original/springer-static/esm/art:10.1038\%2Fs41592-026-03029-6/MediaObjects/41592_2026_3029_MOESM1_ESM.pdf}{\textsuperscript{S}} & Zou Group, Stanford \citep{inheritedC121} & \textit{Runnable analysis application producing Jupyter notebooks}.\newline \href{https://media.springernature.com/original/springer-static/esm/art:10.1038\%2Fs41592-026-03029-6/MediaObjects/41592_2026_3029_MOESM1_ESM.pdf}{Review and revise notebook analyses. Reported human feedback requested donor-level pseudobulk analysis in place of a cell-level comparison.} & \href{https://www.biorxiv.org/content/10.1101/2025.06.03.657517v1}{2025} & \href{https://github.com/zou-group/CellVoyager}{Repository} \\[3pt]
Dango & \citet{proposalDango} & \textit{Web application; public source}.\newline \href{https://arxiv.org/html/2503.03154v2}{Demonstrate table edits, answer clarification questions, and revise generated transformation steps before applying them.} & \href{https://arxiv.org/abs/2503.03154}{2025} & \href{https://ferranschen.github.io/site/dango/dango.html}{Project} \\[3pt]
Data Agent (Li et al., 2026) & \citet{proposalDataAgent} & \textit{Research data system; public software unverified}.\newline \href{https://arxiv.org/html/2609.24137v1}{Validate low-confidence analytical results after the system exhausts internal correction attempts.} & \href{https://arxiv.org/html/2609.24137v1}{2026} & \href{https://arxiv.org/abs/2609.24137}{Paper only} \\[3pt]
Data Formulator & Microsoft Research \citeyearpar{proposalDataFormulator} & \textit{Public analysis and visualization application}.\newline \href{https://github.com/microsoft/data-formulator/blob/5477f0e236426dc8f74a498ec400414fba7fbc0f/README.md}{Choose data-loading alternatives and iteratively revise or branch visual analyses.} & \href{https://arxiv.org/abs/2309.10094}{2023} & \href{https://github.com/microsoft/data-formulator}{Repository} \\[3pt]
Data Interpreter (MetaGPT application) & \citet{hong2024} & \textit{Runnable command-line research scripts and notebook/Python agent}.\newline \href{https://github.com/FoundationAgents/MetaGPT/blob/11cdf466d042aece04fc6cfd13b28e1a70341b1f/metagpt/strategy/planner.py}{In optional manual mode, review plans and task results; request revisions before continuing.} & \href{https://github.com/FoundationAgents/MetaGPT/releases/tag/v0.7.0}{2024} & \href{https://github.com/FoundationAgents/MetaGPT/tree/main/metagpt/roles/di}{Repository} \\[3pt]
data-to-paper\href{https://github.com/Technion-Kishony-lab/data-to-paper/blob/81df14c4b9600466e645c3b2b336cc54daa3df3a/src/data_to_paper/base_steps/request_code.py}{\textsuperscript{S}} & Kishony Lab, Technion \citep{inheritedC101} & \textit{Application GUI and runnable research workflow}.\newline \href{https://github.com/Technion-Kishony-lab/data-to-paper/blob/81df14c4b9600466e645c3b2b336cc54daa3df3a/src/data_to_paper/base_steps/request_code.py}{In optional Copilot mode, review or edit statistical code-review feedback, then send it back for correction of the analysis.} & \href{https://arxiv.org/abs/2404.17605}{2024} & \href{https://github.com/Technion-Kishony-lab/data-to-paper}{Repository} \\[3pt]
Databox AI Analyst & Databox \citeyearpar{proposalDatabox} & \textit{Hosted business analytics; skills/routines}.\newline \href{https://databox.com/rethinking-analytics-agentic-era}{Steer analyses through chat and revise metric definitions that govern subsequent queries.} & \href{https://databox.com/rethinking-analytics-agentic-era}{2026} & \href{https://databox.com/rethinking-analytics-agentic-era}{Update} \\[3pt]
DataFoundry & datagallery-lab (team) \citeyearpar{proposalDataFoundry} & \textit{Public analysis workbench; web/TUI}.\newline \href{https://github.com/datagallery-lab/datafoundry/blob/5cf7d364e2eddf69f84a1ac2eb0a606e069c67a4/docs/en/guides/web-workbench.md}{Review steps, select evidence, and branch alternative analyses from persisted checkpoints.} & n.v. & \href{https://github.com/datagallery-lab/datafoundry}{Repository} \\[3pt]
DATAGEN & \citet{inheritedC085} & \textit{Runnable research workflow with human-review stages}.\newline \href{https://github.com/zi-yue-1129/DATAGEN/blob/main/src/core/workflow.py}{Accept or regenerate hypotheses, then request additional analysis during review.} & n.v. & \href{https://github.com/zi-yue-1129/DATAGEN}{Repository} \\[3pt]
DeepAnalyze / DA-Studio & \citet{proposalDeepAnalyze} & \textit{Public model and analysis application}.\newline \href{https://github.com/ruc-datalab/DeepAnalyze/blob/d4dd5cf62ec7ec2579eb85bcab4f24e4782827b9/demo/chat_v2/README.md}{Optional Manual mode pauses after code execution for feedback; users can revise and rerun scripts in Code Lab.} & \href{https://arxiv.org/abs/2510.16872}{2025} & \href{https://github.com/ruc-datalab/DeepAnalyze}{Repository} \\[3pt]
Deepnote Agent & Deepnote \citeyearpar{proposalDeepnote} & \textit{Hosted notebook agent; paid plans}.\newline \href{https://deepnote.com/docs/deepnote-agent}{Co-edit executable notebooks, inspect changes, and continue from revised notebook state.} & \href{https://deepnote.com/changelog/2025-08-28}{2025} & \href{https://deepnote.com/docs/deepnote-agent}{Docs} \\[3pt]
DSH Data Agent & omdsh-dev (maintainers) \citeyearpar{proposalDSHDataAgent} & \textit{Analysis application on DeepSeek Harness; web/TUI}.\newline \href{https://github.com/omdsh-dev/dsh-data-agent/blob/6a649fe930aeab4247d3e8b936b6bf393aa71e8c/README.md}{Confirm and revise metric definitions reused in subsequent analytical conversations.} & \href{https://github.com/omdsh-dev/dsh-data-agent/releases/tag/v0.0.6}{2026} & \href{https://github.com/omdsh-dev/dsh-data-agent}{Repository} \\[3pt]
Genie Code & Databricks \citeyearpar{proposalGenieCode} & \textit{Agent in Databricks workspace}.\newline \href{https://www.databricks.com/blog/introducing-genie-code}{Collaborate on notebook code and review or roll back analytical workspace changes.} & \href{https://www.databricks.com/blog/introducing-databricks-assistant}{2023} & \href{https://docs.databricks.com/aws/en/genie-code}{Docs} \\[3pt]
Genie One / Genie Agents & Databricks \citeyearpar{proposalGenieAgents} & \textit{Hosted conversational analytics}.\newline \href{https://docs.databricks.com/aws/en/genie-agents/talk-to-genie}{Refine conversations and edit query parameters or curated SQL that shape subsequent results.} & \href{https://www.databricks.com/blog/introducing-aibi-intelligent-analytics-real-world-data}{2024} & \href{https://docs.databricks.com/aws/en/genie-agents/}{Docs} \\[3pt]
GenoMAS & \citet{inheritedNbio14} & \textit{Runnable multi-agent research workflow}.\newline \href{https://github.com/Liu-Hy/GenoMAS/blob/8c2a549a0eb58cff4e09675897c88d8dbbbc6254/README.md}{Optionally review and edit generated scientific Action Units before the computational analysis proceeds.} & \href{https://arxiv.org/abs/2507.21035}{2025} & \href{https://github.com/Liu-Hy/GenoMAS}{Repository} \\[3pt]
Hex Agent / Threads & Hex Technologies \citeyearpar{proposalHex} & \textit{Hosted analytical workspace}.\newline \href{https://learn.hex.tech/changelog/page/7}{Refine analytical plans and co-edit executable notebook cells and visualizations.} & \href{https://learn.hex.tech/changelog/page/7}{2025} & \href{https://hex.tech/}{Product} \\[3pt]
Julius & Julius AI \citeyearpar{proposalJulius} & \textit{Hosted analysis application}.\newline \href{https://julius.ai/docs/learning-subagent/viewing-learned-information}{Follow up conversationally and correct learned schema meanings used in later analyses.} & \href{https://julius.ai/changelog}{2023} & \href{https://julius.ai/}{Product} \\[3pt]
Jupyter AI Agents & Datalayer, Inc. \citeyearpar{proposalJupyterAIAgents} & \textit{Analysis agent; JupyterLab chat UI and CLI}.\newline \href{https://github.com/datalayer/jupyter-ai-agents}{Co-edit and execute notebook cells through a JupyterLab agent interface.} & \href{https://pypi.org/project/jupyter-ai-agent/}{2024} & \href{https://github.com/datalayer/jupyter-ai-agents}{Repository} \\[3pt]
LAMBDA & \citet{sun2025lambda} & \textit{Web application}.\newline \href{https://github.com/AMA-CMFAI/LAMBDA/blob/main/README.md}{Preserved v1 supports editing analysis code and executing it within the continuing agent workflow.} & \href{https://arxiv.org/abs/2407.17535}{2024} & \href{https://github.com/AMA-CMFAI/LAMBDA}{Repository} \\[3pt]
MLE-agent & \citet{inheritedNstatisticsmleagent} & \textit{Interactive command-line application}.\newline \href{https://github.com/MLSysOps/MLE-agent/blob/main/mle/agents/planner.py}{Revise the ML plan and code through explicit stage-level feedback.} & \href{https://github.com/MLSysOps/MLE-agent/blob/main/README.md}{2024} & \href{https://github.com/MLSysOps/MLE-agent}{Repository} \\[3pt]
MUSE\href{https://arxiv.org/html/2608.16181v1}{\textsuperscript{S}} & \citet{chen2026muse} & \textit{Research web application; public software unverified}.\newline \href{https://arxiv.org/html/2608.16181v1}{Inspect warnings about questionable data treatment or evaluation, select repairs, and propagate them through downstream analysis.} & \href{https://arxiv.org/html/2608.16181v1}{2026} & \href{https://arxiv.org/abs/2608.16181}{Paper only} \\[3pt]
PantheonOS & Qiu Lab, Stanford \citep{inheritedC110} & \textit{Agent application with team runtime and user interface}.\newline \href{https://github.com/aristoteleo/PantheonOS/releases/tag/v0.4.9}{Review plans and provide structured feedback before consequential analytical work continues.} & \href{https://www.biorxiv.org/content/10.64898/2026.02.26.707870v1}{2026} & \href{https://github.com/aristoteleo/PantheonOS}{Repository} \\[3pt]
Posit Assistant\href{https://opensource.posit.co/blog/2026-10-09_ai-newsletter/}{\textsuperscript{S}} & Posit Software, PBC \citeyearpar{proposalPosit} & \textit{IDE and terminal data-science agent}.\newline \href{https://opensource.posit.co/blog/2026-10-09_ai-newsletter/}{Review cleaning decisions before the final cleaning script is generated; a demonstration retains extreme flight delays as genuine observations.} & \href{https://posit.co/blog/2026-03-27-ai-newsletter}{2026} & \href{https://assistant.posit.co/}{Docs} \\[3pt]
ProCARE & \citet{proposalProCARE} & \textit{React research web application; public source}.\newline \href{https://github.com/ZyFanTJ/ProCARE/blob/c013e50c64c9edf24335c432c7aad20879788f41/code/frontend/src/pages/Wizard.tsx}{Revise a generated study plan used in subsequent analyses of existing observational health data.} & \href{https://arxiv.org/abs/2610.16175}{2026} & \href{https://github.com/ZyFanTJ/ProCARE}{Repository} \\[3pt]
QwenPaw-Data & \citet{proposalQwenPawData} & \textit{CLI/engine and native data workspace; public source}.\newline \href{https://github.com/agentscope-ai/QwenPaw-Data/blob/e0c8ce54b3a60bc26df11c75a43766dbf8c8a7c2/packages/qwenpaw-data-cli/src/qwenpaw_data/cli/commands/execute.py}{Revise executable plans; the reviewed interaction policy requests confirmation of unresolved metric meanings before SQL generation. Questioning depends on the interface; direct CLI mode disables it.} & \href{https://arxiv.org/abs/2607.11019}{2026} & \href{https://github.com/agentscope-ai/QwenPaw-Data}{Repository} \\[3pt]
RD-Agent & Microsoft \citep{proposalRDAgent} & \textit{CLI and optional application interaction UI}.\newline \href{https://github.com/microsoft/RD-Agent/blob/main/rdagent/log/ui/ds_user_interact.py}{Review candidate hypotheses and edit tasks or instructions before experiments.} & \href{https://pypi.org/project/rdagent/0.0.1/}{2024} & \href{https://github.com/microsoft/RD-Agent}{Repository} \\[3pt]
Sigma Assistant / Agents & Sigma Computing \citep{proposalSigma} & \textit{Hosted analytics and workbooks}.\newline \href{https://www.sigmacomputing.com/blog/announcing-ask-sigma}{Clarify analytical meaning and revise inspectable analysis steps in editable workbooks.} & \href{https://www.sigmacomputing.com/blog/announcing-ask-sigma}{2024} & \href{https://www.sigmacomputing.com/blog/sigma-assistant-upgrade}{Update} \\[3pt]
TexeraAgent & \citet{proposalTexeraAgent} & \textit{Apache Texera workflow canvas and chat; public release}.\newline \href{https://github.com/apache/texera/blob/7349698c55002f7da77992bcdc770cb4d6f49356/agent-service/src/agent/texera-agent.ts}{Edit executable dataflows or attach feedback to past actions, then continue from that version.} & \href{https://texera.apache.org/releases/}{2026} & \href{https://github.com/apache/texera}{Repository} \\[3pt]
Treadstone & \citet{proposalTreadstone} & \textit{Research UI prototype; public software unverified}.\newline \href{https://arxiv.org/html/2609.19774v1}{Curate analytical branches, select suggested directions, and contribute evidence to a shared multi-agent feed.} & \href{https://arxiv.org/html/2609.19774v1}{2026} & \href{https://arxiv.org/abs/2609.19774}{Paper only} \\[3pt]
WebSeek & \citet{huang2026webseek} & \textit{Browser extension and backend; public source}.\newline \href{https://arxiv.org/html/2601.15100v1}{Edit shared tables and charts, accept suggested transformations, and direct subsequent work through chat.} & \href{https://arxiv.org/abs/2601.15100}{2026} & \href{https://github.com/datavisards/WebSeek}{Project} \\[3pt]
\end{longtable}
\endgroup

\section{Selected Applications for Study Preparation Before Data Collection}
\label{app:study-design-systems}
\setcounter{table}{0}
\renewcommand{\thetable}{B.\arabic{table}}
\renewcommand{\theHtable}{B.\arabic{table}}

The applications below support selected tasks before new data are collected, including statistical sizing, experimental authoring, protocol development, and design assessment. These roles and their evaluations differ. A precollection function does not establish prospective validation of the resulting study, and documented participation does not establish the statistical adequacy of a design. Levels and \textsuperscript{S} use the definitions in Appendix~\ref{app:interaction-levels}; these applications are reported separately from its analytical-system counts. OrchestrXR meets the S criterion through its described support for human control of experimental factors. Collecting design parameters or stating that assumptions require human review is insufficient without a documented connection from that review to a design decision.

\begingroup
\footnotesize
\setlength{\tabcolsep}{3pt}
\renewcommand{\arraystretch}{1.12}
\begin{longtable}{@{}L{0.18\textwidth}L{0.20\textwidth}L{0.38\textwidth}L{0.06\textwidth}L{0.10\textwidth}@{}}
\caption{Selected applications for study preparation before data collection. Levels describe documented human participation; \textsuperscript{S} marks documented human interaction guided by statistical principles.}\label{tab:study-design-systems}\\
\toprule
System / level & Developer / authors (reference) & Form / documented participation & Year & Link \\
\midrule
\endfirsthead
\multicolumn{5}{l}{\textit{Table \thetable\ continued}}\\
\toprule
System / level & Developer / authors (reference) & Form / documented participation & Year & Link \\
\midrule
\endhead
\midrule
\multicolumn{5}{r}{\footnotesize Continued on next page}\\
\endfoot
\bottomrule
\endlastfoot
N-Power AI\newline Level 1 & \citet{proposalNPowerAI} & \textit{Web application for sample-size and power analysis}.\newline Submit a study description for automated R-based calculations and a report. & \href{https://www.biorxiv.org/content/10.1101/2025.02.06.636776v1}{2025} & \href{https://ai.swmed.edu/N-PowerAI/}{Product} \\[3pt]
OrchestrXR\href{https://arxiv.org/html/2607.01588v1}{\textsuperscript{S}}\newline Level 3 & \citet{proposalOrchestrXR} & \textit{Paper-described web interface and Unity workflow; public software availability unverified}.\newline Review and selectively revise experimental factors while preserving stable baselines; carry revised specifications into the XR prototype. This is a described design purpose, not established experimental validity. & \href{https://arxiv.org/abs/2607.01588}{2026} & \href{https://arxiv.org/abs/2607.01588}{Paper only} \\[3pt]
TrialAtlas\newline Level 1 & \citet{proposalTrialAtlas} & \textit{Command-line research application}.\newline Submit trial information for autonomous design assessment and proposed improvements. & \href{https://arxiv.org/abs/2609.21859}{2026} & \href{https://github.com/linjc16/trialatlas}{Repository} \\[3pt]
CRISPR-GPT\newline Level 3 & \citet{proposalCRISPRGPT} & \textit{Gene-editing design application}.\newline Supply stepwise design choices and correct user-proxy responses. A light implementation is public; the full implementation is restricted. & \href{https://arxiv.org/abs/2404.18021v1}{2024} & \href{https://github.com/cong-lab/crispr-gpt-pub}{Repository} \\[3pt]
TrialMind\newline Level 3 & Keiji AI \citep{proposalTrialMindDesign} & \textit{Commercial platform}.\newline Review proposed plans and stages involving estimands, power, simulation, and design choices; features are documented by the vendor. & n.v. & \href{https://www.keiji.ai/solutions/biostatistics-trial-design}{Product} \\[3pt]
\end{longtable}
\endgroup

\clearpage
\PaperPrintMainRefsfalse
\begin{paperreferences}{Appendix References}
\addtolength{\bibsep}{-1pt}
\InputIfFileExists{\jobname.bbl}{}{}
\end{paperreferences}

\end{document}